\documentclass[preprint,review,11pt]{elsarticle}

\usepackage[margin=2.5cm]{geometry}
\usepackage{amssymb, amsfonts, subfigure}
\usepackage{amsmath, amsthm}
\usepackage{graphicx}
\newtheorem{thm}{Theorem}
\newtheorem{lm}{Lemma}

\newcommand{\RR}{\mathbb{R}}

\newcommand{\bs}{\boldsymbol}

\journal{Journal of Chinese Statistical Association}

\begin{document}

\begin{frontmatter}



\title{Variable Screenings in Binary Response Regressions with Multivariate Normal Predictors}


\author{Sheng-Mao Chang}

\address{Department of Statistics, National Cheng Kung University}

\begin{abstract}
Screening before model building is a reasonable strategy to reduce the dimension of regression problems. Sure independence screening is an efficient approach to this purpose. It applies the slope estimate of a simple linear regression as a surrogate measure of the association between the response and the predictor 
so that the final model can be built by those predictors with steep slopes. However, if the response is truly affected by a nontrivial linear combination of some predictors then the simple linear regression model is a misspecified model. In this work, we investigate the performance of the sure independence screening in the view of model misspecification for binary response regressions. Both maximum likelihood screening and least square screening are studied with the assumption that predictors follow multivariate normal distribution and the true and the working link function belong to a class of scale mixtures of normal distributions. 
\end{abstract}


\end{frontmatter}


\section{Introduction}\label{sec:intro}
A common characteristic of massive data sets whose major purpose of study is to discover the association between response and predictors is that the number of predictors is larger than the number of independent individuals. Although linear regression or its generalizations are useful tools to detect associations, some computational and theoretical issues are still remained unsolved for massive data. Many statistical approaches have been developed during the past two decades in many aspects. In this work, we concentrate on the screening problem for binary response regressions. Considering all predictors in one linear model is not practical without placing restrictions on the parameter space. Instead, one may design screening statistics to rule out unimportant predictors before building final models. A screening statistic is defined as a surrogate measure of the underlying association between the response and a predictor. Fan and Lv (2008) and Fan and Song (2010) propose the concept of ``sure screening": a screening statistic possesses the sure screening property if that the statistic is relatively small if the true association is negligible or 0. Their sure independence screenings were designed toward this end.

Fan and Lv (2008) proposes the sure independence screening (SIS) for linear regression. They choose the predictors with large absolute covariances between response and predictors as important predictors and then build final regression models based on these important predictors. As we will show later, the covariance is proportional to the slope of the simple linear regression so hereafter, we use slope instead of covariance if there is no confusion. The computation of SIS is extraordinarily fast because it only involves centering and inner products but no matrix inversion. Note that when the term ``linear regression" is applied, we generally presume that the response is continuous or more restrictively, the response follows normal distribution. Either way, least-square estimation can be applied to estimate the regression coefficients.  One of our major interest is the consequence of applying least-square estimation to binary response data. As we will show later, under some conditions, it is useful for screening but not for estimation and prediction.  

For binary response regressions, it is popular to choose logit, probit or complementary log-log link function (McCullagh and Nelder, 1989) to formulate the likelihood. Many statistical softwares perform estimation and testing tasks well. However, there are two major issues on using these models. First, the choice of link function is essential to estimation science different link functions yield different regression coefficient estimates. Li and Duan (1989) proves that, the maximum likelihood estimate (MLE) is consistent to the true regression coefficient up to an unknown constant when the link function is misspecified. So, when the true link function is unknown, the regression coefficient estimates is always questionable. Second, the MLE of regression coefficient is sometimes unidentifiable, unique or finite MLE does not exist (Albert and Anderson, 1984). These two reasons urge us to find a computational efficient procedure for screening rather than merely using traditional logistic or probit regressions.   

The rest of this article is arranged as follows. In Section 2, we review some useful results of linear model as well as the sure independence screening in linear model (Fan and Lv, 2008) and in generalized linear regression (Fan and Song, 2010). Moreover, we show that, for binary response regressions, the screening statistics of both SIS and the newly proposed least-square screening (LeSS) converge in probability to its linear model counterpart up to a constant when the predictors follow multivariate normal and the link function belongs to a class of scale mixture of normals. Simulation and data analysis are shown in Section 4 followed by our concluding remarks in Section 5. 

\section{Regression Model Misspecification}
\subsection{Gaussian Response}
We begin with matching parameters of the true model and parameters of working models. Consider the linear model
\begin{equation} \label{eq:TrueModel}
	Y = \gamma_0 + \sum_{j=1}^p X_j \gamma_j + \varepsilon
\end{equation}
where $\varepsilon \sim N(0,\tau^2)$, $(X_1,\cdots, X_p)$ follows $MVN({\bf 0}, \bs \Sigma)$, and $\varepsilon$ and $X_j$'s are independent.  Assume that not all of $\gamma$'s are 0. Denote ${\bf X}^T = (X_1,\cdots,X_p)$ and $\sigma_{ij}$ as the $(i,j)$ element of $\bs\Sigma$. Let (\ref{eq:TrueModel}) be the true model and call the predictors with non-zero (zero) regression coefficients as active (inactive) predictors. As taught in the first course of linear regression, the least-square estimator of the slop of the working model $E(Y| X_1,\cdots, X_p)= \beta_{0j} + \beta_j X_j$ converges in probability to
\[
	\beta_j = \frac{Cov(Y, X_j)}{Var(X_j)} 
	= \frac{Cov(\sum_{l=1}^p X_l\gamma_l + \varepsilon_1, X_j)}{Var(X_j)}
	=\sum_{l=1}^p \sigma_{lj}\gamma_l/\sigma_{jj} \quad \mbox{for }j=1,\cdots, p.
\]
and hence ${\bs \beta} = {\bf V}^{-1}{\bs \Sigma}{\bs \gamma}$ where ${\bs\beta}^T=(\beta_1,\cdots,\beta_p)$, ${\bs\gamma}^T=(\gamma_1,\cdots,\gamma_p)$, and $\bf V$ is a diagonal matrix with diagonal elements $(\sigma_{11},\cdots, \sigma_{pp})$. 

For a more general case, let the working model be 
\begin{equation}\label{eq:Working1}
	E(Y|X_1,\cdots,X_p)=\beta_0 + \sum_{j=1}^{p_1} X_j\beta_j
\end{equation}	
where $p_1 < p$. Define ${\bf Z}_1^T=(X_1,\cdots,X_{p_1})$ and ${\bf Z}_2^T=(X_{p_1+1},\cdots,X_p)$ and partition the regression coefficient vector as ${\bs\gamma}^T=({\bs\gamma_1}^T, {\bs\gamma}_2^T)$ where ${\bs\gamma}_1$ and ${\bs\gamma}_2$ are regression coefficients corresponding to ${\bf Z}_1$ and ${\bf Z}_2$, respectively. Further, define the partition of the variance covariance matrix as
\[
	{\bs\Sigma} = \left[\begin{array}{cc} 
		{\bs\Sigma}_{11} & {\bs\Sigma}_{12} \\ {\bs\Sigma}_{21} & {\bs\Sigma}_{22} 
	\end{array} \right]
\]
with respect to ${\bf Z}_1$ and ${\bf Z}_2$, too. Then we have $E({\bf Z}_1Y)={\bs\Sigma}_{11}{\bs\gamma}_1 + {\bs\Sigma}_{12}{\bs\gamma}_2$ which implies that the least-square estimate of the working model (\ref{eq:Working1}) converges in probability to 
\begin{equation} \label{eq:main}
	{\bs\gamma}_1 + {\bs\Sigma}_{11}^{-1}{\bs\Sigma}_{12}{\bs\gamma}_2
\end{equation}
This suggests that ${\bs \beta}_1^T=(\beta_1,\cdots, \beta_{p_1})={\bs\gamma}^T_1$ if either ${\bs\Sigma}_{12}={\bf 0}$ or ${\bs\gamma}_2={\bf 0}$ which is actually the partial orthogonality condition defined in Huang, Horowitz and Ma (2008). In other words, to successfully estimate regression coefficients (${\bs\gamma_1}$) without contamination (${\bs\Sigma}_{11}^{-1}{\bs\Sigma}_{12}{\bs\gamma}_2$), a subset of predictors, say ${\bf Z}_1$, should be chosen so that ${\bf Z}_1$ and ${\bf Z}_2$ are uncorrelated or none of predictors in ${\bf Z}_2$ is active. Note that, the multiple regression with $p_1$ predictors is a misspecified model for the true model (\ref{eq:TrueModel}) as long as these $p_1$ predictors do not include all active predictors. A well-known result in linear model literature is that if the working model is misspecified, the least-square estimator of the regression coefficient is biased. The asymptotic bias can be quantified explicitly by (\ref{eq:main}). However, to our best knowledge, there is no such expression under binary response regressions.

\subsection{Binary Response}
The score equation is applied to link the true model parameters and the maximum likelihood estimates of the parameter under a specific misspecified model. Suppose the true model is $Y|{\bf X} \sim Ber(\pi_T)$ and $g(\pi_T) = \gamma_0 + \sum_{j=1}^p X_j\gamma_j$, where $g^{-1}(x) = 1/(1+e^{-x})$ is the so called logit link. Assume that  the true link is known and a simple working model $g(\pi_W)= \beta_0 + \beta_1 X_1$ is specified. Consequently, the score function converges in probability to $E(X_1(Y-\pi_W))$ and thus, the maximum likelihood estimator of $(\beta_0, \beta_1)$ converges to $(\beta_0^{ML}, \beta_1^{ML})$ such that
\[
	0 = E\left[ X_1(Y-\pi_W)\right]=
	E\left[ X_1 \frac{1}{1+\exp\{-\gamma_0 - \sum_{j=1}^p X_j\gamma_j\}} \right]
	- E\left[X_1 \frac{1}{1+\exp\{-\beta_0^{ML} - \beta_1^{ML} X_1\}} \right]
\]
So the relationship between the true model and working model can be quantified by these two expectations. The calculation of these expectations are not trivial and their numerical evaluations had been studies by Crouch and Spiegelman (1990) and Monahan and Stefanski (1992). One of the major contribution is providing a closed-form expression of these expectations. It is worth to emphasize that $\gamma_1 \neq \beta^{ML}_1$ in general and we wish to express $\beta_1^{ML}$ in terms of true parameter values $\gamma_j$'s like (\ref{eq:main}). Hereafter, denote $\phi(\cdot)$ and $\Phi(\cdot)$ as the probability density function ({\em p.d.f.}) and cumulative distribution function ({\em c.d.f.}) of the standard normal distribution, respectively, and denote $\phi(\cdot; \mu, \tau^2)$ as the density function of the normal distribution with mean $\mu$ and variance $\tau^2$. 

Now, we derive the relationship between the regression coefficient of the true model and of the working model for binary response regression models.  Let the true model be
\begin{equation}\label{eq:BTrue}
	Y|X_1,\cdots, X_p \sim Ber(\pi_T), \quad \pi_T 
	= H_T\left(\gamma_0 + \sum_{j=1}^p X_j\gamma_j\right)
\end{equation}
and the working model be
\begin{equation}\label{eq:BWorking}
	Y|X_1,\cdots, X_p \sim Ber(\pi_W), \quad \pi_W 
	= H_W\left(\beta_0 + \sum_{j=1}^{p_1} X_j\beta_j\right)
\end{equation}
where a function with the subscript $W$ means that the function is unknown but one is posited to it, and the function with subscript $T$ means that the function is the underlying function. Moreover, we require that $H_W(\cdot)$ as well as $H_T(\cdot)$ is a valid {\em c.d.f.} with the form of scale normal mixture
\[
	H_W(t) = \int_{\RR^+} \Phi(t/\sigma)q_W(\sigma) d\sigma
\]
and $q_W(\sigma)$ is a valid density function for either continuous or discrete $\sigma$. This implies that $h_W(t) = dH_W(t)/dt$ is a symmetric {\em p.d.f.} around 0. Such an $h_W(\cdot)$ can be the {\em p.d.f.} of Gaussian, logistic, double exponential, Student-$t$ (Andrews and Mallows, 1974), exponential power family (Box and Tiao, 1973; West, 1987) and others. Hereafter, $H_W(t)$'s should satisfy 1) $H_W(t) = \int_{\RR^+} \Phi(t/\sigma)q_W(\sigma) d\sigma$, 2) $q_W(\sigma)$ is a valid density function, and 3) $\int_{\RR} \phi(t;\mu,\sigma^2+\tau^2)q(\sigma)d\sigma < \infty$ for every $|\mu| < \infty$ and $\tau^2 < \infty$. A sufficient and necessary condition for the existence of $q_W(\sigma)$ is provided by Andrews and Mallows (1974). Following is one of our major conclusion and it is the result of Lemma \ref{lm:AB00} and Lemma \ref{lm:probit}. Theirs proofs are deferred to Appendix A. Theorem \ref{th:main} implies that the least square estimator $\tilde {\bs\beta}_1$ converges to a value proportional to the desire value (\ref{eq:main}) and the proportion is expressed in a form of integration. 
\begin{thm}\label{th:main}
	Under the true model (\ref{eq:BTrue}) and the working model (\ref{eq:BWorking}), the least-square estimator $\tilde{\bs\beta}_1$ converges in probability to ${\bs\beta}_1^{LS} = {\bs\Sigma}_{11}^{-1}Cov({\bf Z}_1, Y)
		 = \left({\bs\gamma}_1 + {\bs\Sigma}_{11}^{-1}{\bs\Sigma}_{12}{\bs\gamma}_2\right)
		 \times c_1$
	where $c_1=\int_{\RR^+}\phi(0; \gamma_0, \sigma^2+{\bs\gamma}^T{\bs\Sigma}{\bs\gamma}) q_T(\sigma)d\sigma$.
\end{thm}

\begin{lm} \label{lm:AB00} Arnold and Beaver (2000) prove that
\begin{enumerate}
	\item Linearly skewed normal ${\bf X} \sim LSN(\lambda_0,{\bs\lambda}_1)$ is defined according to the following equation
		\[
			\int_{\RR^{p}} \phi({\bf x})\Phi(\lambda_0 + {\bs\lambda}_1^T {\bf x}) d{\bf x} 
			= \Phi\left( \frac{\lambda_0}{\sqrt{1+{\bs\lambda}_1^T{\bs\lambda}_1}} \right)
		\]
	\item If ${\bf X} \sim LSN(\lambda_0,{\bs\lambda}_1)$ then 
			\[
				E({\bf X}) = {\bs\lambda}_1\frac{1}{\sqrt{1+{\bs\lambda}_1^T{\bs\lambda}_1}}
				\phi\left(\frac{\lambda_0}{\sqrt{1+{\bs\lambda}_1^T{\bs\lambda}_1}}\right)
				\left[\Phi\left(\frac{\lambda_0}{\sqrt{1+{\bs\lambda}_1^T
					{\bs\lambda}_1}}\right)\right]^{-1}
			\]
\end{enumerate}
\end{lm}

\begin{lm}\label{lm:probit}
Under the true model (\ref{eq:BTrue}) with probit link ($q(\sigma) = I(\sigma=1)$), 
\[
	E({\bf Z}_1Y) = \left({\bs\Sigma}_1{\bs\gamma}_1 + {\bs\Sigma}_{12}{\bs\gamma}_2\right)
	\phi(0; \gamma_0, 1+{\bs\gamma}^T{\bs\Sigma}{\bs\gamma})
\]
\end{lm}

Under the working model (\ref{eq:BWorking}), we show that the maximum likelihood estimator of ${\bs\beta}_1$ converges in probability to ${\bs\beta}_1^{ML} \propto {\bs\beta}_1^{LS}$ in Theorem \ref{th:MLE} where the proof is rooted from the score equation 
$0 = E\left\{{\bf Z}_1 \left[Y-H_W({\bf Z}^T_1{\bs\beta}^{ML}_1)\right] \right\}
	= E\left\{{\bf Z}_1\left[H_T({\bf Z}^T{\bs\gamma})-H_W({\bf Z}^T_1{\bs\beta}^{ML}_1)\right]\right\}$  
which implies that the theorem holds according to Theorem \ref{th:main} and we omit. Note that Theorem \ref{th:MLE} also says that when $H_T \neq H_W$, the link function is misspecified, the regression coefficient estimator of the working model converges to a value which is proportional to the true regression coefficient. This conclusion is also made in Li and Duan (1989) but they do not provide an explicit form of the proportion. 

\begin{thm}\label{th:MLE}
	Under the true model (\ref{eq:BTrue}) and the working model (\ref{eq:BWorking}), the maximum likelihood estimator $\hat{\bs\beta}_1$ converges in probability to  ${\bs\beta}_1^{ML} = {\bs\beta}_1^{LS} / c_2(\beta_0^{ML},{\bs\beta}_1^{ML})$ where $c_2(\beta_0^{ML},{\bs\beta}_1^{ML})=\int_{\RR^+}\phi(0; \beta^{ML}_0, \sigma^2+({\bs\beta}_1^{ML})^T{\bs\Sigma}_{11}{\bs\beta}_1^{ML}) q_W(\sigma)d\sigma$.
\end{thm}

The integrations involved in Theorem \ref{th:main} and Theorem \ref{th:MLE} are not easy to evaluate because $q_T(\sigma)$ and $q_W(\sigma)$ may not have simple forms. For example, Stefanski (1990) shows that when $h_W(t)$ is the density function of the logistic distribution, $q_W(\sigma)=dL(\sigma/2)/d\sigma$ and $L(\sigma) = 1 - 2\sum_{j=1}^{\infty}(-1)^{j+1}\exp\{ -2j^2\sigma^2\}$. A friendly computation form can be
\[	
	\int_{\RR^+}\phi(0; \beta_0, \sigma^2+{\bs\beta}^T{\bs\Sigma}{\bs\beta}) q_W(\sigma)d\sigma 
		= \left\{ \begin{array}{ll}
		h_W(\beta_0) & 
		\mbox{if } {\bs\beta}= {\bf 0} \\
		\int_{\RR} \phi(t;\beta_0, {\bs\beta}^T{\bs\Sigma}{\bs\beta}) h_W(t) dt &
		\mbox{if } {\bs\beta} \neq {\bf 0} \\
	\end{array} \right.
\]
which can be numerically evaluated by many computer softwares easily when all parameters and $h_W$ are specified. Finally, we emphasize that the inverse function of the complementary log-log link is the {\em c.d.f.} of extreme value distribution, and since the corresponding {\em p.d.f.} is not symmetric around 0, Theorem 1 does not apply to this case. 

\section{Variable Screenings}
In this section, we formally define the screening statistics including the SIS and the LeSS. For the $i$th individual, $i=1, \cdots, n$, let the true model be
\[
	Y_i|X_{i1},\cdots, X_{ip} \sim Ber(\pi_i) \quad \pi_i 
	= H_T\left(\gamma_0 + \sum_{j=1}^p x_{ij}\gamma_j\right)
\]
and the working model for the $k$th screening statistic be
\[
	Y_i|X_{i1},\cdots, X_{ip} \sim Ber(\pi_i) \quad \pi_i 
	= H_W\left(\beta_0 + X_{ik}\beta_k\right)
\]
Denote the likelihood of the working model as $l(\beta_k, H_W)$. Similarly, define the least-square objective function as $Q(\beta_k) = \sum_{i=1}^n (Y_i-\beta_0 - X_{ik}\beta_k)^2$. For the SIS (Fan and Song, 2010), the screening statistic is $\hat \beta_k = \mbox{argmax } l(\beta_k, H_W)$. We further distinguish it into two methods: SIS with probit link (SISP), $\hat \beta^P_j = \mbox{argmax } l(\beta_k, \Phi)$, and SIS with logit link (SISL), $\hat \beta^L_j = \mbox{argmax } l(\beta_k, L)$ where $L(t)=1/(1+e^{-t})$. Moreover, the screening statistic of LeSS is $\tilde\beta^{LS}_k = \mbox{argmax } Q(\beta_k)$. In addition, because $\beta_0$ plays no role in screening so we omit but readers should know that above maximizations are taken with respect to $\beta_0$ and $\beta_k$ simultaneously. By Theorems \ref{th:main} and \ref{th:MLE}, we conclude that $\tilde\beta_j^{LS} \rightarrow \beta_j^{LS}$, $\hat\beta_j \rightarrow \beta_j^{ML}$, and $\beta^{ML}_j \propto \beta^{LS}_j \propto \gamma_j + e_j$ where $e_j=\sum_{i \neq j} \sigma_{ij} \gamma_i / \sigma_{jj}$ is the contamination due to other active predictors. If $e_j\approx 0$ then $\gamma_j=0$ implies $\beta_j^{ML} \approx 0$. A general condition for $e_j\approx 0$ is that both $\bs\gamma$ and $\bs\Sigma$ are sparse (Fan and Lv, 2008). Finally, the SIS recommends choosing $d$ largest $|\hat\beta^{L}_k|$'s to build final models where $d= n / \log n$ and $n$ is the sample size. We shall follow this rule for SISP and LeSS procedures.

\begin{figure}[!ht]
\begin{center}
	\includegraphics[height=4in,width=5.5in]{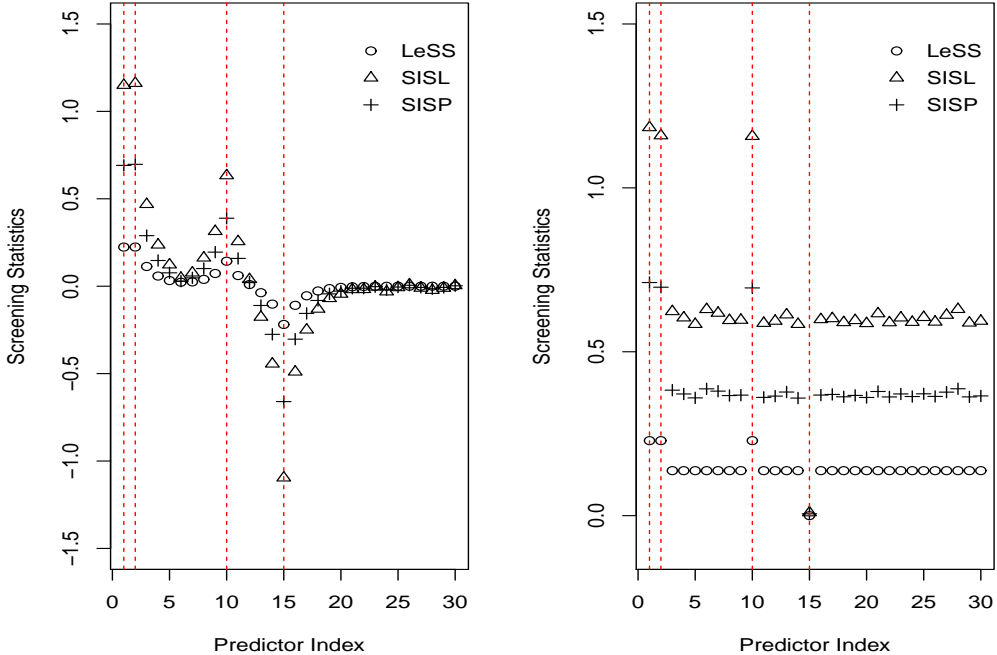}
\end{center}
\caption{Averages of screening statistics: The dotted lines indicate the indices of active predictors. The left panel shows the result of AR1 model whereas the right panel shows the result of CS model.}
\label{fig:ASS}
\end{figure}

Following (true) model is designed to demonstrate the consequence of model misspecification. Suppose that predictors $(X_1,\cdots, X_{30})$ follows normal distribution with zero mean and the true covariance structure is either autoregression (AR1), $\sigma_{ij}=\rho^{|i-j|}$, or compound symmetry (CS), $\sigma_{ij}=I(i=j)+\rho I(i\neq j)$ where $I(A)=1$ if $A$ is true and $I(A)=0$ otherwise. Let the true model be $Y\sim Ber(\pi)$ and $\Phi^{-1}(\pi) = X_1 + X_2 + X_{10} - 3\rho X_{15}$
where $\rho=0.5$. By Theorem \ref{th:main}, $\beta_k^{LS}$'s can be evaluated according to the true model. Unfortunately, there is no closed-form expression for $\beta_k^{ML}$'s and therefore, we estimate $\beta^{ML}_k$'s by averaging 100 maximum likelihood estimates where each of them was calculated based on 200 independent samples drawn from the true model. The results are shown in Figure \ref{fig:ASS}. This figure expresses the intuition that the sure screening property does not hold for all kind of data sets. From Figure \ref{fig:ASS}, when the predictors have the AR1 correlation structure (the left panel), both LeSS and SIS seems to have high chance to detect all active predictors. On the other hand, when predictors have the CS correlation structure (the right panel), the predictor $X_{15}$ is barely chosen by all mentioned screening method.

\section{Numerical Evidences}
\subsection{Simulation Study}
A simple scheme was applied to demonstrate Theorem 1. Let the true model be $Y|X_1,\cdots, X_5 \sim Bin(1, \pi)$ and $\pi=E(Y|X_1,\cdots, X_5) = H_T(X_1 + X_2 - 2X_3)$
where the link function is either the probit link or the logit link. Moreover, we generate normal predictors $(x_1,\cdots, x_{5})$ with zero mean and aforementioned covariance structures AR1 and CS. For each dataset, 200 {\em i.i.d.} samples were generated and 100 datasets were simulated. Table \ref{tb:bias} shows that if the link function and the true regression coefficients are known then the  least-square regression coefficient estimates with adjustment ($c_1 \times \tilde\beta_k$'s) seem to be consistent. 
\begin{table}
\caption{Biases of Adjusted Least Square Estimation ($\rho=0.5$)}
\label{tb:bias}
\begin{center}
\begin{tabular}{llrrrrrrr}
  \hline\hline
  Cov. & Link & Stat. & $\beta_1$ & $\beta_2$ & $\beta_3$ & $\beta_4$ & $\beta_5$ & $c_1$\\
  \hline
  AR1 & probit & mean & -0.000 & -0.010 & -0.013 & 0.021 & -0.046 & 0.178\\ 
  && s.e. & 0.029 & 0.033 & 0.030 & 0.031 & 0.028 \\ 
  & logit & mean & 0.018 & 0.013 & -0.004 & 0.027 & -0.029 & 0.151\\ 
  && s.e. & 0.040 & 0.043 & 0.036 & 0.037 & 0.037 \\ 
  CS & probit & mean & -0.003 & -0.027 & -0.005 & 0.007 & 0.010 & 0.199\\ 
  && s.e. & 0.026 & 0.032 & 0.026 & 0.030 & 0.025 \\ 
  & logit & mean & 0.048 & -0.053 & 0.022 & -0.048 & 0.019 & 0.164\\ 
  && s.e. & 0.035 & 0.033 & 0.034 & 0.035 & 0.037 \\ 
  \hline\hline
\end{tabular}
  $(\beta_1,\beta_2,\beta_3,\beta_4,\beta_5)=(1,1,-2,0,0)$
\end{center}
\end{table}

Next, we investigated the performance of screening statistics. The simulation plan is as follows. For each dataset, $1000$ predictors with $n$ independent samples were generated with the true model $Y|X_1,\cdots, X_p \sim Bin(1, \pi)$ and $\pi= E(Y|X_1,\cdots, X_p) = \Phi\left( X_1 + X_2 + X_{10} - 3\rho X_{15} \right)$ 
This model is analogous to the model applied in Simulated example III of Fan and Lv (2008). It is designed so that the response and $X_{15}$ are uncorrelated under Gauss-Markov model and that the covariance structure among predictors is the CS structure. The true covariance structure among these 1000 predictors is AR1 or CS with $\rho=0.5$ and the underlying link function is the probit link. We tried  $n=100$, $200$, and $500$. So following the suggestion of Fan and Lv (2008), the number of chosen predictors $(d)$ are 21, 37, and 80 ($d=n/\log n$), respectively. Additionally, we are also interested in the performance of these screening methods under the circumstance that predictors does not follow normal distribution. To this end, we generated correlated binomial predictors sequentially by the approach proposed by Biswas and Hwuang (2002). The data generating process is shown in Appendix B. The predictors were generated so that $X_i \sim Bin(2, \pi_i)$ where $\pi_i \in (0.1, 0.5)$, random; correlations between two consecutive predictors are also random. Screening results for both normal and binomial data are summarized in Table \ref{tb:acu}. The performance of a screening method is evaluated by the rate of choosing all active predictors at the same time for a given $d$.

\begin{table}
\caption{Rates of choosing all active predictors}
\label{tb:acu}
\begin{center}
\begin{tabular}{l|ccc|ccc|ccc}
  \hline\hline
	Cov & \multicolumn{3}{c|}{Normal-AR1} & \multicolumn{3}{c|}{Normal-CS} 
	& \multicolumn{3}{c}{Correlated Binomial}\\ 
	\hline 
	$n$ & 100 & 200 & 500 & 100 & 200 & 500 & 100 & 200 & 500\\
	\hline
	SISL & 0.63 & 0.97 & 1.00 & 0.00 & 0.00 & 0.01 & 0.16 & 0.76 & 1.00 \\
	SISP & 0.63 & 0.97 & 1.00 & 0.00 & 0.00 & 0.01 & 0.16 & 0.73 & 1.00 \\
	LeSS & 0.63 & 0.97 & 1.00 & 0.00 & 0.00 & 0.01 & 0.08 & 0.64 & 1.00 \\ 
  \hline\hline
\end{tabular}
\end{center}
\end{table}

From Table \ref{tb:acu}, we observe that three screening methods have the same rate of choosing all active predictors for normally distributed predictors. However, when the predictors are sampled sequentially and correlated from binomial distributions, SISL and SISP have higher rates than LeSS. As shown in Figure \ref{fig:ASS}, under the CS covariance structure, the screening approaches are inappropriate and hence the 15th predictor has a very low chance to be selected. For other simulation setup, not surprisingly, when sample size increases, all LeSS, SISL, and SISP have higher rates of choosing all active predictors at the same time. 

\subsection{Variable Screening: discriminating two leukemia types}
In this study, 7128 gene expression levels for 72 patients were recorded. Among these patients, 45 patients was diagnosed as acute lymphoblastic leukemia and the other 27 patients was diagnosed as acute myeloid leukemia. The dataset is available on the website {\it http://statistics.standarford.edu/$\sim$brad}. The goal of the analysis is to identify putative genes which can distinguish two specific leukemia types. To this end, logistic regression can be applied. However, the total number of predictors (gene expression levels) is too large to include all of them in a single model, and hence we screened these predictors first. Following a particular screening method, a discrimination rule was built by using a logistic regression with screened genes as covariates. Then, we had three discrimination rules associated with SISP, SISL and LeSS, respectively. To examine the performances of these three rules, misclassification rates of discrimination were considered. For each screening approach, 16 ($\approx 72/\log 72$) out of 7218 genes were selected. The selected genes were tabulated in Table \ref{tb:genes}. There are 15 common selected genes between SISL and SISP and only 8 common selected genes among SISL, SISP, and LeSS. The misclassification rates for SISP, SISL, and LeSS, are 24/72, 25/72, and 24/72, respectively. However, when the 8 common genes are included in a single model, the misclassification rate is 0. 

\begin{table}
\caption{Ranking of screened genes}
\label{tb:genes}
\begin{center}
\begin{tabular}{r|rrrrrrrr}
  \hline\hline
  Ranking & 1 & 2 & 3 & 4 & 5 & 6 & 7 & 8 \\ \hline
  SISL & 4847 & 1962 & {\bf 6041} & {\bf 2121} & {\bf 1882} & {\bf 1834} & 2128 & {\bf 804} \\ 
  SISP & 4847 & 1962 & {\bf 2121} & {\bf 6041} & {\bf 1834} & 2128 & {\bf 1882} & {\bf 804} \\ 
  LeSS & 3252 & 2288 & {\bf 1834} & {\bf 6854} & 760 & {\bf 1882} & {\bf 804} & 2354 \\ 
	\hline\hline
  Ranking & 9 & 10 & 11 & 12 & 13 & 14 & 15 & 16 \\ \hline
  SISL & {\bf 6854} & 2020 & 2111 & 4366 & {\bf 1144} & {\bf 1745} & 4196 & 2402 \\ 
  SISP & {\bf 6854} & 4366 & 2020 & 2111 & {\bf 1144} & 2402 & {\bf 1745} & 4377 \\ 
  LeSS & 5501 & {\bf 1144} & {\bf 2121} & 1685 & {\bf 1745} & 4328 & 4211 & {\bf 6041} \\ 
   \hline\hline
\end{tabular}
	\\
	Bold-faced numbers are gene IDs that discovered by all of three screening methods.
\end{center}
\end{table}
	
\section{Concluding Remarks}
In this work, we investigate the screening statistics, SISL, SISP, and LeSS, in population level. The quality of screening depends on the sparseness of the true regression coefficient and the sparseness of the covariance structure of predictors. This conclusion can be drawn from (\ref{eq:main}) and numerically proved by our simulations. Also, none of these methods satisfies the sure screening property defined by Fan and Lv (2008) without placing some restrictions on these two sparseness. Fortunately, in many experiments, it is nature to make these sparseness assumptions. The leukemia dataset demonstrated in Section 4.2 is a good example. First, it is reasonable to assume that only a few genes affects the subtypes of leukemia and second, indeed, the covariance among gene expressions is low on average. Moreover, Dudoit {\em et al.} (2002) analyze this dataset via many classification methods. Among these approaches, the diagonal linear discriminate analysis, assuming that all gene expressions are uncorrelated, beats others. This may imply that the sparseness of the covariance structure assumption and the Gaussian assumption on gene expressions fit the observed data well. 

Although both the SIS's and the LeSS are defined in a very restrictive manner, we suggest two rules to apply them. First, if the covariance structure is rich then aforementioned screening approaches would fail. Under this circumstance, using other delicate statistical approaches is recommended. Second, if predictors do not follow normal distribution then the sample size should be large for an acceptable screening result. When there is no significant violation of above conditions and the dataset is huge, we advocate using LeSS because it is very computational efficient. For the leukemia dataset, SISL and SISP cost 23 and 31 seconds, respectively, whereas LeSS cost less than 1 second.

\section*{Appendix A}
Proof of Theorem \ref{th:main}
\begin{proof}
Note that the least square estimator converges in probability to the inverse of the covariance matrix among predictors multiplied by the covariance  between the response and the predictors. The former is trivial and thus we show the latter. We begin with the conditional expectation 
\[
	E({\bf Z}_1 Y|{\bf Z}_2={\bf z}_2) = E({\bf Z}_1 H({\bs\gamma}_1^T{\bf X}_1 + C_1)|{\bf Z}_2={\bf z}_2)
	= \int_{\RR^{p_1}} {\bf z}_1 H(C_1 + {\bs\gamma}_1^T{\bf z}_1)
	\phi({\bf z}_1; {\bs\mu}, {\bs\Omega}) d {\bf z}_1
\]
where $C_1 = \gamma_0 + \sum_{j=p_1+1}^p X_j\gamma_j$, ${\bs\mu} = {\bs\Sigma}_{12}{\bs\Sigma}_{22}^{-1} {\bf z}_{2}$ and ${\bs\Omega} = {\bs\Sigma}_{11} - {\bs\Sigma}_{12}{\bs\Sigma}_{22}^{-1}{\bs\Sigma}_{21}$.
According to the proof of Lemma 2,
\[\begin{split}
	E({\bf Z}_1Y|{\bf Z}_2={\bf z}_2) &=\int_{\RR^{p_1}} {\bf z}_1 
	\left[
		\int_{\RR^+} \Phi\left(\frac{C_1 + {\bs\gamma}_1^T{\bf z}_1}{\sigma}\right)
		q(\sigma)d\sigma
	\right]	
	\phi({\bf z}_1; {\bs\mu}, {\bs\Omega}) d {\bf z}_1\\
	&=\int_{\RR^+}
	\left[
		\int_{\RR} {\bf z}_1  \Phi\left(\frac{C_1 + {\bs\gamma}_1^T{\bf z}_1}{\sigma}\right)
		\phi({\bf z}_1; {\bs\mu}, {\bs\Omega}) d {\bf z}_1
	\right]	
	q(\sigma)d\sigma
\end{split}\]
and therefore 
\[
	E({\bf Z}_1Y) = \int_{\RR^+}({\bs\Sigma}_{11}\tilde{\bs\gamma}_1 + {\bs\Sigma}_{12}\tilde{\bs\gamma}_2) 
	\phi(0;\tilde\gamma_0, 1 + \tilde{\bs\gamma}^T{\bs\Sigma}\tilde{\bs\gamma})
	q(\sigma)d\sigma
\]
where $\tilde\gamma_0=\gamma_0/\sigma$, $\tilde{\bs\gamma}_1={\bs\gamma}_1/\sigma$, $\tilde{\bs\gamma}_2={\bs\gamma}_2/\sigma_2$, and $\tilde{\bs\gamma}={\bs\gamma}/\sigma$. After change of variable, we have
\[
	E({\bf Z}_1Y) = ({\bs\Sigma}_{11}{\bs\gamma}_1 + {\bs\Sigma}_{12}{\bs\gamma}_2) \int_{\RR^+} 
	\phi(0;\gamma_0, \sigma^2+{\bs\gamma}^T{\bs\Sigma}{\bs\gamma}).
	q(\sigma)d\sigma
\]
Last, since $\int_{\RR^+}\phi(0;\gamma_0, \sigma^2 + {\bs\gamma}^T{\bs\Sigma}{\bs\gamma})q(\sigma)d\sigma$ exists, $E({\bf Z}_1Y)$ exists.
\end{proof}

Proof of Lemma \ref{lm:probit}
\begin{proof}
Define $C_1 = \gamma_0 + \sum_{j=p_1+1}^p x_{1j}\gamma_j$. Then
\[
	E({\bf Z}_1Y|{\bf Z}_2={\bf z}_2) = E({\bf Z}_1 H({\bs\gamma}_1^T{\bf Z}_1 + C_1)|{\bf Z}_2={\bf z}_2)
	= \int_{\RR^{p_1}} {\bf z}_1 \Phi(C_1 + {\bs\gamma}_1^T{\bf z}_1)\phi({\bf z}_1; {\bs\mu}, {\bs\Omega}) 
	d {\bf z}_1
\]
where ${\bs\mu} = {\bs\Sigma}_{12}{\bs\Sigma}_{22}^{-1}{\bf z}_{2}$ and ${\bs\Omega} = {\bs\Sigma}_{11} - {\bs\Sigma}_{12}{\bs\Sigma}_{22}^{-1}{\bs\Sigma}_{21}$. Let ${\bf u} = {\bs\Omega}^{-1/2}({\bf z}_1-{\bs\mu})$. The conditional expectation $E({\bf Z}_1 Y|{\bf Z}_2={\bf z}_2)$ can be rewritten as
\[
	\int_{\RR^{p_1}} ({\bs\mu} + {\bs\Omega}^{1/2} {\bf u}) 
	\Phi(C_1 + {\bs\gamma}_1^T{\bs\mu} + {\bs\gamma}_1^T{\bs\Omega}^{1/2} {\bf u})
	\phi({\bf u}) d {\bf u}.
\]
Further, let $\lambda_0 = C_1 + {\bs\gamma}_1^T{\bs\mu}$ and ${\bs\lambda}_1 = {\bs\Omega}^{1/2}{\bs\gamma}_1$. By Lemma \ref{lm:AB00}, above integration becomes
\[
	{\bs\mu}\Phi\left( \frac{\lambda_0}{\sqrt{1+{\bs \lambda}_1^T{\bs\lambda}_1}}\right) 
		+ {\bs\Omega}{\bs\gamma}_1\frac{1}{\sqrt{1+ {\bs\lambda}_1^T{\bs\lambda}}}
		\phi\left( \frac{C_1 + {\bs\gamma}^T_1{\bs\mu}}{\sqrt{1+ {\bs\lambda}_1^T{\bs\lambda}}} \right)
\]
or equivalently
\[
	{\bs\Sigma}_{12}{\bs\Sigma}_{22}^{-1} {\bf Z}_2 
		\Phi\left( \frac{\gamma_0 + {\bf b}^T{\bf Z}_2}{\sqrt{1
		+ {\bs\gamma}_1^T{\bs\Omega}{\bs\gamma}_1}}\right)
		+ {\bs\Omega}{\bs\gamma_1}\phi\left( {\bf b}^T{\bf Z}_2 ; - \gamma_0, 
		1+ {\bs\gamma}_1^T{\bs\Omega}{\bs\gamma}_1\right)
\]
where ${\bf b} = {\bs\Sigma}_{22}^{-1}{\bs\Sigma}_{21}{\bs \gamma}_1 + {\bs\gamma}_2$. Since ${\bf Z}_2 \sim N({\bf 0}, {\bs\Sigma}_{22})$ and $w= {\bf b}^T{\bf z}_2 \sim N(0, {\bf b}^T{\Sigma}_{22}{\bf b})$, 
\[\begin{split}
	E({\bf Z}_1Y) =& {\bs\Sigma}_{12}{\bs\Sigma}_{22}^{-1} E\left[ {\bf Z}_2 
		\Phi\left( \frac{\gamma_0 + {\bf b}^T{\bf Z}_2}{\sqrt{1
		+ {\bs\gamma}_1^T{\bs\Omega}{\bs\gamma}_1}}\right)\right]\\
		&+ {\bs\Omega}{\bs\gamma_1} \int_{\RR} \phi\left( w ; - \gamma_0, 
		1+ {\bs\gamma}_1^T{\bs\Omega}{\bs\gamma}_1\right) \phi(w; 0, {\bf b}^T{\bs\Sigma}_{22}{\bf b}) dw.
\end{split}\]
Again, by Lemma \ref{lm:AB00}, 
\[
	E({\bf Z}_1Y) =({\bs\Sigma}_{12}{\bf b} + {\bs\Omega}{\bs\gamma}_1) \phi(0;\gamma_0, 
	1 + {\bs\gamma}_1^T{\bs\Omega}{\bs\gamma}_1 + {\bf b}^T{\bs\Sigma}_{22}{\bf b})
\]
and therefore, after some algebra, $E({\bf Z}_1Y) =({\bs\Sigma}_{11}{\bs\gamma}_1 + {\bs\Sigma}_{12}{\bs\gamma}_2)\phi(0; \gamma_0, {\bs\gamma}^T{\bs\Sigma}{\bs\gamma})$.
\end{proof}


\section*{Appendix B}
Let $(X_1, X_2)$ follow bivariate binomial distribution with marginal distribution $X_1 \sim Bin(2, p_1)$ and $X_2 \sim Bin(2, p_2)$. Following Biswas and Hwang (2002), the joint distribution can be $Pr(X_i=x_1, X_2=x_2) = Pr(X_1=x_1) \times f_{X_2|X_1=x_1}(x_2; p_1, p_2, \alpha)$ where
\begin{equation}\label{eq:cond}
	f_{X_2|X_1=x_1}(x_2; p_1, p_2, \alpha) = \left\{
		\begin{array}{ll}
		{2 \choose x_2} \tilde p_1^{x_2} (1-\tilde p_1)^{2-x_2} & \mbox{ if } x_1=0, \; x_2=0,1,2\\
		(1-\tilde p_1)(1-\tilde p_2) & \mbox{ if } x_1=1, \; x_2=0\\
		(1-\tilde p_1)\tilde p_2 + \tilde p_1 (1-\tilde p_2) & \mbox{ if } x_1=1, \; x_2=1\\
		\tilde p_1 \tilde p_2 & \mbox{ if } x_1=1, \; x_2=2\\
		{2 \choose x_2} \tilde p_2^{x_2} (1-\tilde p_2)^{2-x_2} & \mbox{ if } x_1=2, \; x_2=0,1,2\\
	\end{array}\right.,
\end{equation}
$\tilde p_1 = [p_2+\alpha (p_2-p_1)]/(1+\alpha)$, $\tilde p_2 = \tilde p_1 + \alpha/(1+\alpha)$, and $\alpha$ is a carefully chosen constant. Note that $f_{X_2|X_1}$ is not necessarily a probability mass function for arbitrary $(p_1, p_2, \alpha)$. A sufficient condition to make $f_{X_2|X_1=x_1}$ a probability mass function is that
\begin{equation} \label{eq:suf}
	\frac{\alpha}{1+\alpha} p_1 \leq p_2 \leq \frac{\alpha}{1+\alpha}p_1 + \frac{1}{1+\alpha}.
\end{equation}
Moreover, the correlation coefficient between $X_1$ and $X_2$ is
\[
	\frac{\alpha}{1+\alpha} \sqrt{\frac{p_1(1-p_1)}{p_2(1-p_2)}}
\]
given by Biswas and Hwang (2002). Consequently, an algorithm to simulate correlated $Bin(2, p_i)$ is as follows:
\begin{enumerate}
	\item Sample $q_1$ from $U(0.1, 0.5)$. Sample $x_1$ from $Bin(2, q_1)$. Set $j=2$.
	\item For simulating $x_j$, sample $q_j$ from $U(0.1, 0.5)$ and $\alpha_j$ from $U(0.5, 1)$.
	\item Given $(p_1, p_2, \alpha)=(q_{j-1}, q_j, \alpha_j)$, if the sufficient condition (\ref{eq:suf}) is not satisfied, let $\alpha_j=0$. 
	\item Sample $x_j$ from  the conditional probability (\ref{eq:cond}) with $(p_1, p_2, \alpha)=(q_{j-1}, q_j, \alpha_j)$.
	\item Let $j=j+1$. Goto 2. while $j \leq p$ and stop while $j>p$.
\end{enumerate}
Note that, setting $\alpha=0$ makes two binomial random variables independent. So Step 3. enforces two consecutive variables to be independent with probability roughly equal to 0.1 according to simulation. Moreover, the resulting correlation ranges from 0.2 to 0.8 with median 0.4 by simulation. 
\end{document}